\documentclass[aps,pra,floats,12pt]{revtex4}
\usepackage{graphicx}
\usepackage{amsmath}
\usepackage{amssymb}
\usepackage{color}
\usepackage{ulem}

\renewcommand{\a}{{\alpha}}
\newcommand{\s}{{\sigma}}
\newcommand{\w}{{\omega}}
\renewcommand{\th}{{\theta}}
\renewcommand{\t}{{\tau}}
\renewcommand{\l}{{\lambda}}

\newcommand{\D}{{\Delta}}
\renewcommand{\d}{{\delta}}
\newcommand{\mum}{{$\mu$m}}
\newcommand{\cmm}{{cm$^{-1}$}}

\newcommand{\tnear}{{\tau_{near}}}
\newcommand{\tfar}{{\tau_{far}}}

\newcommand{\beq}{\begin{equation}}
\newcommand{\eeq}{\end{equation}}
\newcommand{\bea}{\begin{eqnarray}}
\newcommand{\eea}{\end{eqnarray}}

\begin{document}
\title{Prospects of coherent control in turbid media:
 Bounds on focusing {{broadband}} laser pulses. }

\begin{abstract} We study the prospects of controlling
transmission of {{broadband and bi-chromatic}} laser pulses through turbid  
samples. The ability to focus transmitted broadband light is
limited via both the scattering properties of the medium, and the
technical characteristics of the experimental set-up. There are
two time scales, given by pulse stretching in the near- and
far-field regions, which define the maximum bandwidth of a pulse
amenable to focusing.
In the {{geometric optics regime of wave propagation in the
medium}}, a single set-up can be optimal for focusing light at
frequencies $\w$ and $n\w$ simultaneously, providing the basis for
the $1+n$ coherent quantum control. Beyond the regime of
{{geometric optics}}, we discuss a simple solution for the
shaping, which provides the figure of merit for one's ability to
focus simultaneously several transmission modes.
\end{abstract}

\date{\today}
\author{Evgeny A.~Shapiro$^1$, Thomas M. Drane$^1$,  Valery
Milner$^{2}$,~\\\it Departments of Chemistry$^1$ and Physics$^2$,
The University of British Columbia
\\  2036 Main Mall, Vancouver, BC, Canada V6T 1Z1}

\maketitle

\section{Introduction.}

This work is motivated by the goal of applying quantum coherent
control techniques in turbid samples. One of the basic ideas of
quantum control involves focusing laser fields of the frequencies
$\w$ and $n \w$ onto an object with nonlinear response.
Interference of the excitation pathways due to each frequency
component creates an asymmetric excitation
\cite{ShapiroBrumerBook}. This effect has been demonstrated in a
number of theoretical and experimental works devoted to generating
directed currents, control of absorbtion and propagation of light,
and breakup processes in various physical and chemical systems
\cite{ShapiroBrumerBook,CohControl-JPB08,Dantus-Review,Silberberg-Review,CohControl-1}.
Extensions of this principle, based on applying ultrafast laser
pulses with controlled broad spectrum, have lead to {{numerous}}
applications in control of quantum evolution, quantum information
processing, spectral characterization, detection, microscopy and
manipulations with microscopic and nano-scopic objects
\cite{ShapiroBrumerBook,CohControl-JPB08,Dantus-Review,Silberberg-Review,Rabitz,RiceBook,ShapedControl}.
We are interested in both the ''$1+n$'' scenario and control with
shaped ultrafast pulses. This task requires an ability to focus
either bi-chromatic or broadband laser pulses with shaped spectrum
in space in time.

As a laser pulse is applied to a turbid sample -- such as ground
glass, biological tissue, paint, suspension, plastic, etc -- its
temporal and spatial structure breaks down
\cite{Ishimaru-Book,Tatarski-book,Goodman-book,Lagendijk-review,Genack-review}.
In space, a coherent beam breaks into a multitude of speckles, so
that spatial focusing is destroyed. In the spectral domain, the
spectrum at each point in space can be strongly modified, so that
the pulse shape is destroyed. The two effects are related, and
each of them is deleterious for coherent control.

This paper analyzes control of transmission of light with multiple
frequency components in turbid samples, with the goal of designing
quantum control experiments. For narrowband light, the
corresponding technique \cite{Mosk-OptLett07} has recently lead to
a breakthrough in focusing and manipulating laser beams in opaque
samples
\cite{Cizmar-NPhot10,Gigan-TMeasure-PRL10,Mosk-NPhot10,Mosk-PRL08}.
The method is based on using a two-dimensional phase mask for the
spatial correction of the wave front.
 We analyze the capabilities of this
approach for spatio-temporal shaping of ultrafast laser pulses.
While the first tests have demonstrated the great potential of the
method for temporal focusing
\cite{Chatel-Focusing-11,Lagendijk-Focusing-11,Silberberg-11},
efficiency of control over the broad bandwidth of ultrafast pulses
needs to be thoroughly understood. Indeed, an experimental set-up
optimized for controlling transmission at one given frequency, may
not be suitable for another \cite{Mosk-BBfocusing-OL11}. A set-up
built to focus light at many frequencies simultaneously may be far
from optimal for each individual spectral component. This work
questions the fundamental limits of controlling broadband
transmission through an opaque sample
\cite{Cizmar-NPhot10,Gigan-TMeasure-PRL10,Mosk-NPhot10,Mosk-PRL08,Lagendijk-Focusing-11,Silberberg-11,Mosk-BBfocusing-OL11}.

We find that the the ability to focus transmitted broadband light
is limited via both the scattering properties of the medium, and
the characteristics of the Spatial Light Modulator (SLM) used to
modify the incident wave front. There are two time scales, given
by pulse stretching in the near- and far-field regions (defined
further in the text), which set the upper limit of the bandwidth
of a pulse that can be focused. Their consideration suggests an
optimization of the experimental set-up. In the {{geometric optics
regime of wave propagation inside the sample}}, a single set-up
can be optimal for focusing light at frequencies $\w$ and $n\w$
simultaneously, providing the basis for the $1+n$ coherent quantum
control, as demonstrated by our
numerical simulations. %
 Beyond geometric optics, i.e. when multiple
interference can not be neglected, there is a simple figure of
merit for one's ability to focus simultaneously several
transmission modes in space. We also discuss a potential ability
of using an opaque sample for shaping broadband spectrum,
effectively replacing the dispersion element in the conventional
pulse shaper \cite{Silberberg-11,Lagendijk-Focusing-11}.


The rest of the paper is organized as follows. In the next Section
we describe the implied experimental set-up, formulate our task in
details, and describe the numerical simulations used throughout
the text for illustration purposes. In Section
\ref{Section-Geometric} we neglect dispersion and backscattering,
and solve the problem in the geometric optics regime, where the
typical scatterer size is bigger than the laser wavelength. Thus
we find the bounds on focusing imposed by the finite modulation
depth of the SLM. In Section \ref{Section-MoreOngeometric}, we
extend the description, including the effects of dispersion,
finite spatial resolution of the phase masks's pixels, and of
focusing of a laser pulse in time and at an angle. We also discuss
focusing of broadband pulses in time vs. focusing in space. The
general case, which goes beyond the geometric optics regime, is
considered in Section \ref{Section-GeneralCase}, where we discuss
the scaling of the problem, and a simple strategy for using SLM to
control simultaneously several independent transmission modes. In
the last Section we summarize the findings of this paper.

\section{Set-up.   \label{Section-SetUp}}

\begin{figure}
\centering
\includegraphics[width=0.75\columnwidth]{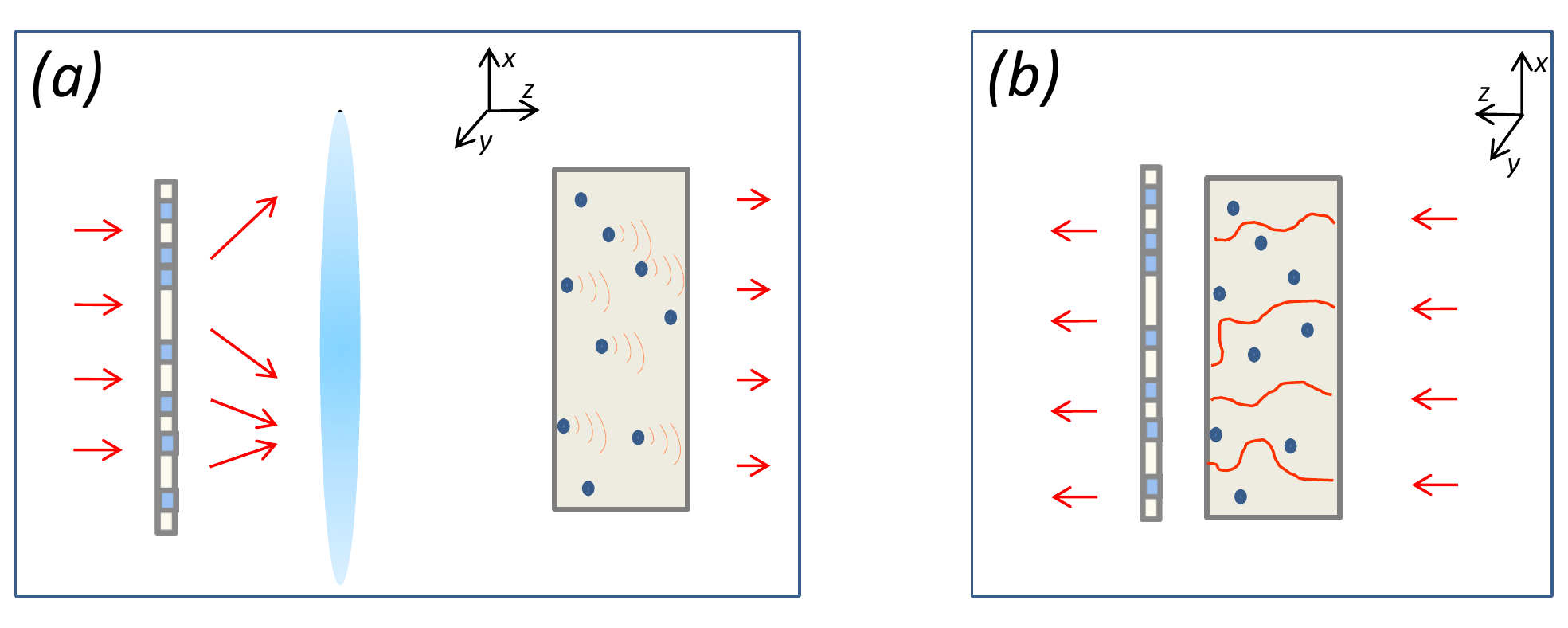}
\caption{(Color online)Implied experimental set-up. In panel (b),
the SLM and imaging lens are replaced by the image of the SLM {{on
the input surface of the sample}}. In addition, the direction of
light propagation is reversed.}
  \vskip -.1truein
\label{Fig-setup}
\end{figure}

Fig.\ref{Fig-setup}(a) shows the general set-up according to Refs.
\cite{Cizmar-NPhot10,Gigan-TMeasure-PRL10,Mosk-NPhot10,Mosk-PRL08,Lagendijk-Focusing-11,Silberberg-11}.
The wavefront of a laser beam is modified by a two-dimensional SLM,
whose pixels add a phase to the incident wavefront. The beam is
then sent onto the scattering sample. Such a configuration allows the optimization of spatio-temporal focusing in
either the near- or far field.

Below, the ``near field region'', $E_{near}(x,y,\w)$, corresponds
to the output surface of this sample.
%
The other, ''far field'' region, with the field distribution
$\tilde E$, is at infinity along the $z$ axis For a spatial
harmonic transmitted at an angle $\theta$,
\begin{equation}
\tilde E(k,\theta) = \int  E_{near}(x,y)\, \exp[i k x \sin\theta]
\,dx\,dy
\label{EfarEnear}%
\end{equation}

In this paper we concentrate on scattering that is sufficiently
treated in the eikonal regime. We limit our consideration to
focusing in the far-field, since it allows for easier modelling.
This would be equivalent to optimizing transmission into a
particular spatial harmonic of $E_{near}$, which can be then
focused with a lens. As we explain below, the temporal structure
of the pulse remains largely undisturbed in the considered regime
and we primarily discuss focusing in space.

Numerically, we solve the scalar wave equation for the electric
field amplitude in the parabolic approximation
\cite{Ishimaru-Book}. The random medium is modelled by a set of
planes. Each plane modifies the wave front as if the light was
passing through a thin glass slide (refractive index $n=1.51$)
with randomly placed "impurities" characterized by a variation $\D
n$ in the refractive index. An example is shown in Fig.2(a). Here
the glass slide is taken to be 10-$\mu$m thick, and round
Gaussian-shaped impurities of the $1/e$ radius $\sigma=30$-$\mu$m
are characterized by $\D n \leq 0.2$. In the calculations, we
place several such planes one after another, separating them by
regions of empty space.

Such modelling is inspired by experiments with diffusors based on
random arrays of  microlenses, ground glass, and all other opaque
materials with relatively large (at least several microns in size)
impurities \cite{Exps-Silberberg,Cui-11,RPCdiffusors}. A single
slide in our modelling creates a far-field speckle pattern, but
does not strongly modify the pulse spectrum. An array of slides,
placed one after another, modifies both the spatial and temporal
structure of a broadband pulse. Although our modelling misses the
effects of de-polarization and backscattering, it allows one to
understand some of the most important aspects of random
propagation in the regime of low to moderate {{scattering angles
(small backscattering)}}. At the same time, the calculations are
fast, allowing us to look at many frequency modes. Numerical
propagation at each frequency consists of applying a
coordinate-dependent phase to the wavefront at the location of
each glass slide, followed by free propagation between
the slides. %
{{The latter is made by making Fourier transform into the wave
vector space and applying a $k$-dependent phase to each spatial
mode.}} For a femtosecond laser pulse sent into the sample, the
temporal shape is obtained as a Fourier transform of the
transmitted spectrum.

\begin{figure}
\centering
\includegraphics[width=\columnwidth]{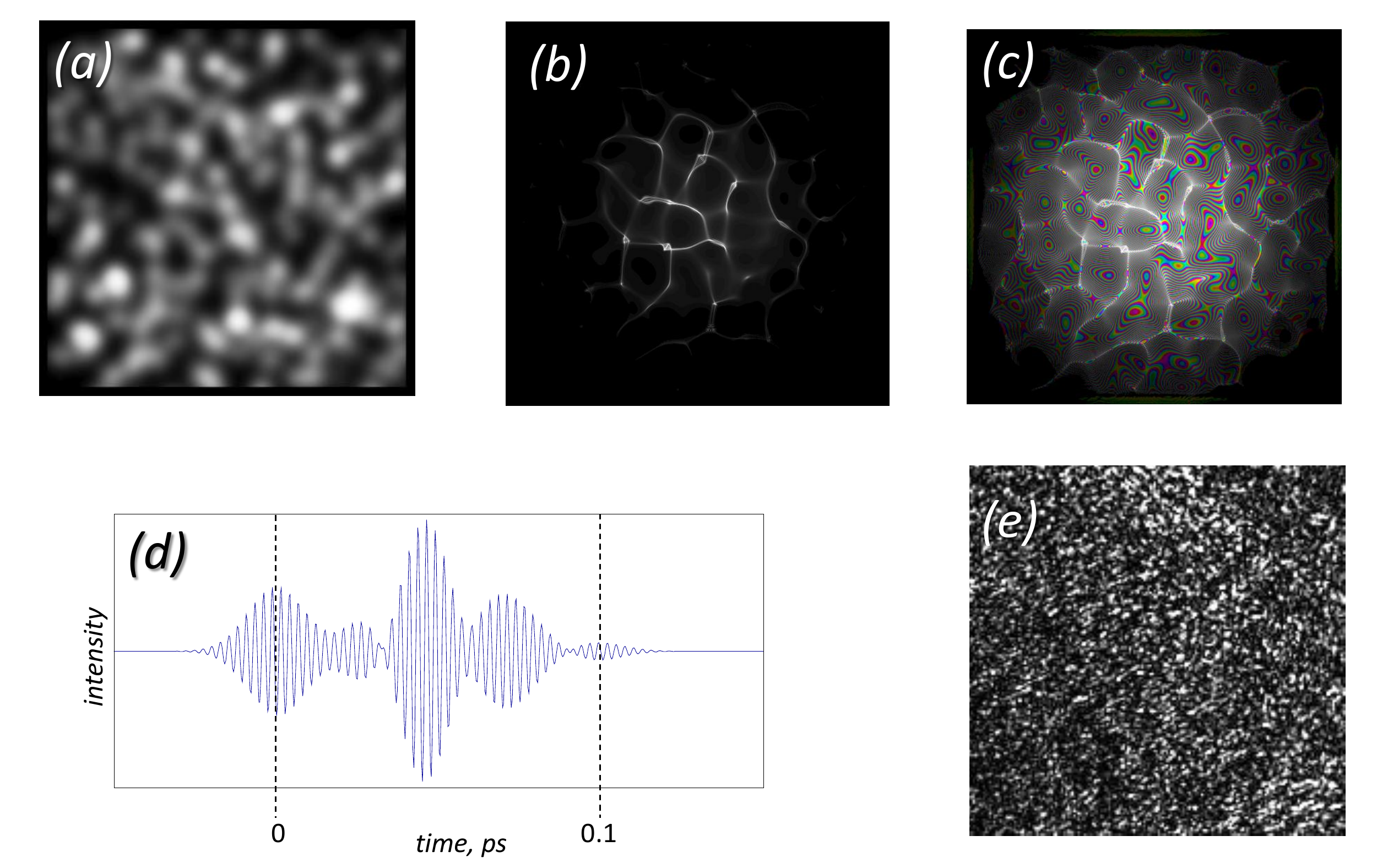}
\caption{ (Color online).(a). Distribution of impurities in one of
the four planes modelled in the calculation. The area shown is
$770\times770$ $\mu$m. (b,c) Near-field intensity and phase
distributions for a $\l_0=800$ nm beam after passing through four
diffusing planes separated by regions of empty space. In panel
(c), intensity is shown by brightness, and phase, between 0 and
2$\pi$, is shown by color. (d) A 25-fs input pulse exits the
system in the far field stretched to 100 fs. (e) Far-field speckle
pattern for 800-nm light.}
  \vskip -.1truein
\label{Fig-Uncompensated}
\end{figure}

Fig.\ref{Fig-Uncompensated}(b) shows the near-field intensity of a
200-\mum -wide beam which has passed through a set of five planes
with $\s=30$ $\mu$m impurities. Adjacent planes are separated by
30 $\mu$m of empty space. In the regime of geometric optics
($\s\gg\l$), the speckle pattern is mainly due to multiple random
lensing. Fig.\ref{Fig-Uncompensated}(c) shows the same beam,
stressing the phase at the exit from the last plane. The phase
pattern is shown for the wavelength $\l=800$ nm. The speckle
pattern at each frequency is almost the same, except for a
frequency dependent phase which corresponds to a different time
delay of the pulse arriving at different $x,y$ points. The zeroth
spatial harmonic is $\tilde E_0 = \int E_{near}(x,y) \,dx\,dy$. A
25-fs pulse sent to the system stretches in the far field to
{{about 100 fs, as shown in Fig.\ref{Fig-Uncompensated}(d).
Fig.\ref{Fig-Uncompensated}(e) shows the far-field speckle pattern
for a $\l=800$ nm beam in the absence of the wave front
compensation.

Here we propose to image the SLM onto the input surface of a
sample. Hereafter, we refer to the image of the SLM as ISLM. In
this geometry, maximizing the transmission from the 0-th to the
0-th spatial mode ($\tau_{00}$)
is achieved simultaneously for forward- and for backward-
propagating beams. Thus each pixel of the phase mask must add to
the {\it backward-propagating} beam an $x,y$-dependent phase such
as to make the wavefront phase as flat as possible
(Fig.\ref{Fig-setup}(b)).

Within the arrangement of Fig.\ref{Fig-setup}(b), we shall use the
term ''near field'' for the field in the ISLM plane -- even if it
is placed at some distance from the actual border of the turbid
sample.

\begin{figure}
\centering
\includegraphics[width=\columnwidth]{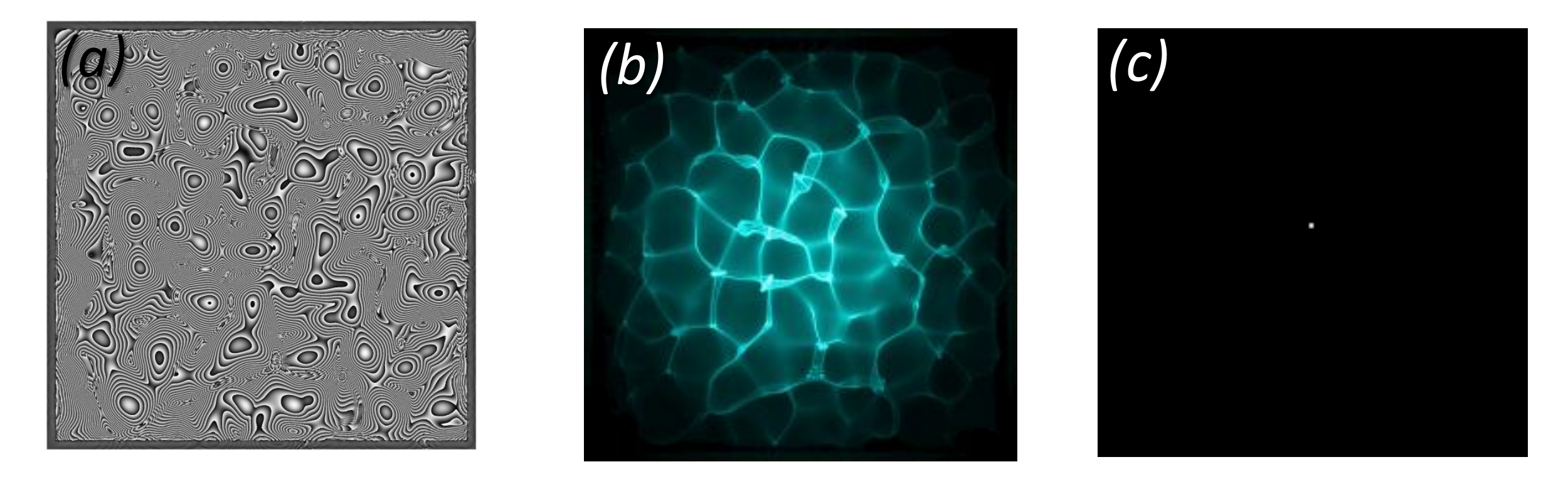}
\caption{(Color online)(a) Phase mask applied by ISLM to correct
the wave front shown in Fig.\ref{Fig-Uncompensated}(c). (b)
Near-field intensity and phase distribution at 800 nm after the
phase correction. (c) Corrected far-field intensity distribution.}
  \vskip -.1truein
\label{Fig-Compensated}
\end{figure}

\section{The role of the SLM's parameters.   \label{Section-Geometric}}

ISLM can be thought of as a thin transparent plate with variable
refractive index $n_{SLM}(x,y)$. A wavefront passing through it
acquires the phase
\beq%
\phi_{SLM}(x,y) = k \a_{0} - k \a (x,y) %
\eeq%
where the optical pathways $\a_0 - \a(x,y)$ are defined by
$n_{SLM}(x,y)=n_0+\D n_{SLM}(x,y)$. Fig.\ref{Fig-Compensated}
shows focusing of the $\l_0=800$ nm wave shown in
Fig.\ref{Fig-Uncompensated}, with an ISLM of infinite spatial
resolution, and a phase modulation depth of $2\pi$ $(0 \leq \a(x,y
\leq \lambda_0) $. Panel (a) shows the mask $\a(x,y)$ ranging from
0 (black) to $\l_0$ (white). Panel (b) shows the corrected
wavefront. Flat phase of the near-field wavefront ensures that the
wave is almost perfectly focused in the far field, as seen in
Panel (c). Variations in the near-field intensity somewhat
decrease the focusing efficiency, adding a broad low-intensity
pedestal, invisible at the scale of Fig.~\ref{Fig-Compensated}(c).

We begin by neglecting dispersion and backscattering, and
considering propagation in the eikonal regime. The latter
corresponds to impurities in the sample being large, $|\nabla n |
\ll k$, where $k$ is the wave vector \cite{Tatarski-book}.
Experiments using commercially available diffusors, ground glass,
random arrays of waveguides, etc, may
 fall under this case. In the eikonal regime, the wave is
composed of trajectories - ''rays''. Each ray propagates in accord
with the laws of geometric optics, and carries the phase $k S$,
where $S$ is the optical path. The surface of equal phase at each
point is orthogonal to the ray passing through this point;
intensity variations are due to the varying density of the rays.

Assume that the SLM's image has sufficient spatial resolution, and
that the SLM is optimized to focus light with the wave vector
$k_0=2\pi/\l_0$. What happens with a wave characterized by
$k=k_0+\D k$? If the maximum ISLM's depth $\a$ was infinite, then
the phase flattening would work perfectly at each frequency.
Indeed, by imaging the SLM mask on the {{surface}} of the sample
we can effectively build a flat slab out of the sample and ISLM:
for each $k$,
\begin{equation}
k\times [\a_{0}+ S_0 + \D S(x,y)- \a(x,y)] = const
 \label{InfiniteAlpha}
\end{equation}
where $S(x,y)=S_0 + \D S(x,y)$ is the optical path of a ray
passing through the point $(x,y)$ at the ISLM's plane.

However, in reality the modulation depth $\a$ can cover only a few
wavelengths. Assuming $\a_{max}=\l_0$, we have for the compensated
wave front at $k_0$
\begin{equation}
k_0\times [\a_{0}+ S_0 + \D S(x,y)- \a(x,y)] = 2\pi j_{xy}+k_0
(\a_0+ S_0)
 \label{FiniteAlpha-k0}
\end{equation}
where $j_{xy}$ is an integer which can vary from one point $(x,y)$
to another. This is the situation shown in
Fig.\ref{Fig-Compensated}. The term $k_0 S_0$ is a constant phase
which does not influence focusing. For a different wave vector, we
have
\begin{equation}
(k_0+ \D k)\times [\a_{0}+S_0 + \D S(x,y) - \a(x,y) ] = %
(k_0+ \D k)(\a_0 + S_0) + 2\pi j_{xy} + \D k (\D S(x,y) - \a(x,y))
~.
 \label{FiniteAlpha-DK}
\end{equation}
The phase compensation (\ref{FiniteAlpha-DK}) will work for any
$k$ if the maximum modulation depth $\a_{max}>\D S$ for most
pathways. If, on the other hand, $\D S \gg \a_{max}$, the
compensation will not work as soon as {{$\D k (\D S - \a) \simeq
\D k \D S$ exceeds $\pi$}} for many points $(x,y)$. Thus
\begin{equation}\D k_{max}= {\pi \over
 \langle \D S \rangle} \label{DkMax} ~.
\end{equation}
According to the Huygens – Fresnel principle, %
\begin{equation}
\langle \D S\rangle = c \,\tfar ~,
\label{Taufar}\end{equation}%
where $\tfar$ shows how much a short pulse sent to the system is
stretched in the 0-th spatial mode or, equivalently, in the
far-field focus. Another way to see this fact is as follows.
Consider two points, A and B, at the exit from the sample, such
that $S_A = S_B + \langle \D S\rangle$. When CW light of the
frequency $\w$ is sent into the system, the phase of the field at
the points A and B differs by $\phi_{AB}(\w) = \langle \D S\rangle
\w /c $. At a different frequency, $\w + \d\w$,
\begin{equation}
\phi_{AB}(\w+\d\w) = \phi_{AB}(\w) + \d\w \langle \D S\rangle / c
\label{DwFar}
\end{equation}
According to Eq.(\ref{EfarEnear}), the complex values of the field
from all near-field points are summed to produce a far-field
speckle.  One can see from Eq.(\ref{DwFar}) that  the detuning
$\d\w = \pi c /\langle \D S\rangle $ corresponds to the speckle
pattern being significantly different from that at the frequency
$\w$. At this value of the detuning, constructive interference
between the fields coming into the far field region from the
points A and B turns into a destructive one, and vice versa.
Therefore, the frequency correlation length of the far field
speckle pattern is, approximately $2\D\w_{far} = 2\pi c /\langle
\D S \rangle$. This means that the transmitted spectrum in the far
field consists of independent bands of the width $2\D\w_{far}$.
Equivalently, a very short laser pulse sent into the system
stretches in the far field to $\t_{far} = \pi / \D\w_{far}$.

Comparing Equations (\ref{DkMax}) and (\ref{Taufar}), we see that
compensation can only work for detunings $\D\w=\w-\w_0$ such that
\begin{equation}
|\D\w|<\frac{\pi}{\tfar} \equiv \D\w_{far} ~.
 \label{FFbound}
\end{equation}
%
\begin{figure}
\centering
\includegraphics[width=0.85\columnwidth]{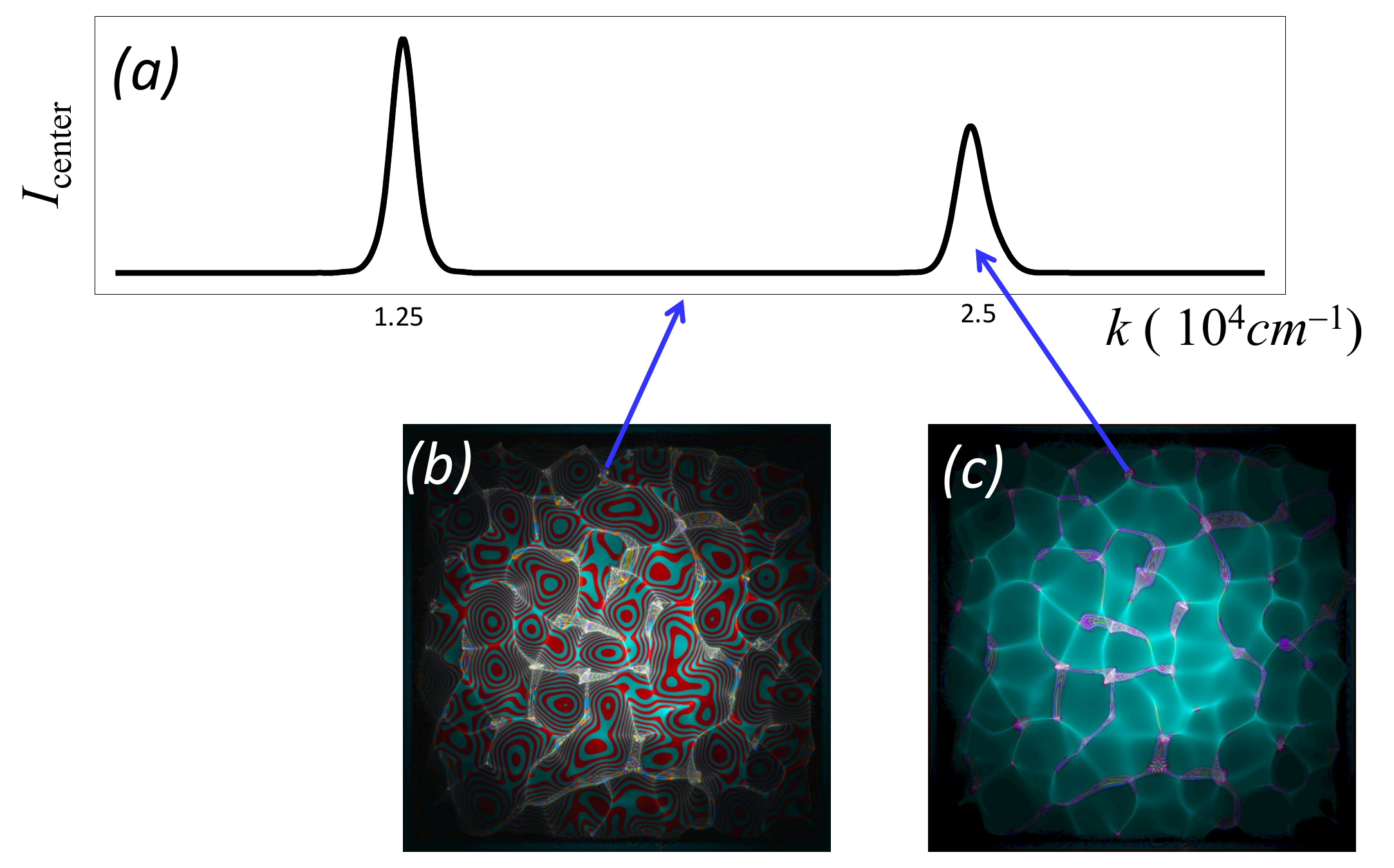}
\caption{(Color online) (a) Laser intensity in the zeroth
far-field mode in dependance on frequency, with the ISLM mask
tuned for $\w_0$. (b,c) Near-field wave fronts at $\w=1.5\w_0$ and
$\w=2\w_0$.}
  \vskip -.1truein
\label{Fig-CompensationScan}
\end{figure}
%
Figure \ref{Fig-CompensationScan} assumes the compensation mask
shown in Fig.\ref{Fig-Compensated}(a) applied to the sample
discussed in Figs.\ref{Fig-Uncompensated} and
\ref{Fig-Compensated}. Panel (a) shows the intensity $I_{center}$
in the  far-field focus in dependance on the field wave vector,
calculated for a single realization of the random sample. As $k$
is detuned from $k_0=12500$ \cmm, the focusing vanishes. {The
width $2\D k \simeq 500$ \cmm corresponds, up to a numerical
factor of $\simeq 1.5$, to $\tfar= 100$ fs}, i.e. the pulse
stretching seen in Fig.\ref{Fig-Uncompensated}(d).

Unexpectedly, $I_{center}$ in Fig.\ref{Fig-CompensationScan}
increases again in the vicinity of $k=2\, k_0$. The effect is
explained in the following way. If the condition
(\ref{FiniteAlpha-k0}) is fulfilled, then
\begin{equation}
2k_0\times [ \a_{0}+S_0 + \D S(x,y) - \a(x,y)] = 4\pi j_{xy}+2k_0
(S_0+S_{0,SLM}) ~,
 \label{FiniteAlpha-2k0}
\end{equation}
and the phase compensation at twice the main wave number is again
complete, as shown in Fig.\ref{Fig-CompensationScan}(c). We see
that in the simplified model --- negligible dispersion, ISLM can
spatially resolve the phase front, --- an ability to focus
bi-chromatic fields, and to perform ''1+$n$'' quantum control
comes at no expense. An experimental set-up optimized to focus a
laser field at frequency $\w$ will also focus field at frequency
$n\w$.

Note that the peak amplitude at  $2 k_0$ in
Fig.\ref{Fig-CompensationScan} is slightly smaller than that at
$k_0$. Indeed, the assumption that the phase mask is able to
resolve individual pathways becomes invalid at the near-field
caustics, where several rays intersect at the same point. This
situation is mathematically similar to that of an SLM with limited
spatial resolution, discussed in the next Section.

Moreover, similar to the case of $k=2k_0$, at $k=1.5\,k_0$ the
phase of the compensated wave can only have two values, 0 and
$\pi$, as seen in Fig.\ref{Fig-CompensationScan}(b). Each part of
the near-field wave front -- that with the zero phase, and the
phase equal to $\pi$ -- yields a strong focus in far-field. The
two foci interfere destructively. However, because of the random
amplitudes, the destructive interference is not complete, and
focusing at $k=1.5\,k_0$ is still better than that at the adjacent
values of $k$. Reminiscent of fractional quantum wave packet
revivals \cite{revivals}, such incomplete focusing happens at any
$k = P/Q \times k_0$ with integer $P,Q$.

\section{Additional bounds.    \label{Section-MoreOngeometric}}

The above consideration remains valid if the goal is to optimize
transmission into a spatial harmonic propagating at an angle $\th$
(or, equivalently, off-axis far field focusing). In this case,
Eq.(\ref{FiniteAlpha-k0}) turns into
\begin{equation}
k_0\times [\a_{0}+ S_0 + \D S(x,y) - \a(x,y)] = 2\pi j_{xy}+k_0
(\a_0+
S_0) + k_0 x\,\sin\th %
\label{FiniteAlpha-k0-angle}
\end{equation}
where $x$ is the coordinate in the ISLM plane. In a compete
analogy with Eq.(\ref{FiniteAlpha-2k0}), light with the wave
vector $2 k_0$ will also be focused. In our numerical simulations,
the spectral bandwidth $\D k$ of the spatially focused light did
not depend on $\th$.

The ability of the scheme to focus several frequencies
simultaneously depends on the sample's dispersion. Indeed, the
above consideration is based on the assumption that light at each
frequency propagates along the same set of rays.  In another
series of calculations we included the effect of dispersion,
assuming that the samples are made of BK-7 glass \cite{BK7}. We
found that the focusing survives in the presence of dispersion: In
our calculations, the focused intensity at the frequency $2\w_0$
decreases only by a factor of, approximately, 2-3. This number is
small compared to the $\sim 10^5$-fold increase in the intensity
at the focus observed in the case of complete phase compensation.

Finite size of the ISLM's pixels in the $(x,y)$ plane does bring
an important additional bound on one's ability to focus broadband
light. If the ISLM grid can not resolve the phase variations in
the incident wave front, then each pixel will be used to
compensate the phase of the average  field
\begin{equation}
E_{av}(k_0) = \sum_S P(\D S_p) e^{i k_0 (S_0 + \D S_p)}
\label{NearFieldSum-k0}
\end{equation}
where $P(\D S_p)$ is the probability distribution for the pathways
characterized by the length $S_0 + \D S_p$ averaged by a single
pixel. Coarse graining over ISLM's pixel size limits the
compensation fidelity. Suppose that a pixel is tuned to compensate
the phase of $E_{av}$ at the given position at the frequency $k_0
c$. At a different frequency we have
\begin{equation}
E_{av}(k) = \sum_S P(\D S_p) e^{i [ k S_0 + k_0 \D S_p + \D k \D
S_p]}~. \label{NearFieldSum-Dk}
\end{equation}
The values of $E_{av}$ correspondent to $k$ and $k+\D k$ differ
drastically if $\D k \D S_p \sim \pi$ for many pathways passing
through the particular pixel. Thus the phase compensation will not
work for detunings $\D\w$ exceeding
\begin{equation}
\D\w_{near} = \frac{\pi}{\tnear}%
\label{DwNear}
\end{equation}
where $\t_{near} = c\, \langle \D S_p \rangle $ describes
stretching of the pulse in the near field, averaged over an area
of the ISLM's pixel.
%
\begin{figure}
\centering
\includegraphics[width=0.8\columnwidth]{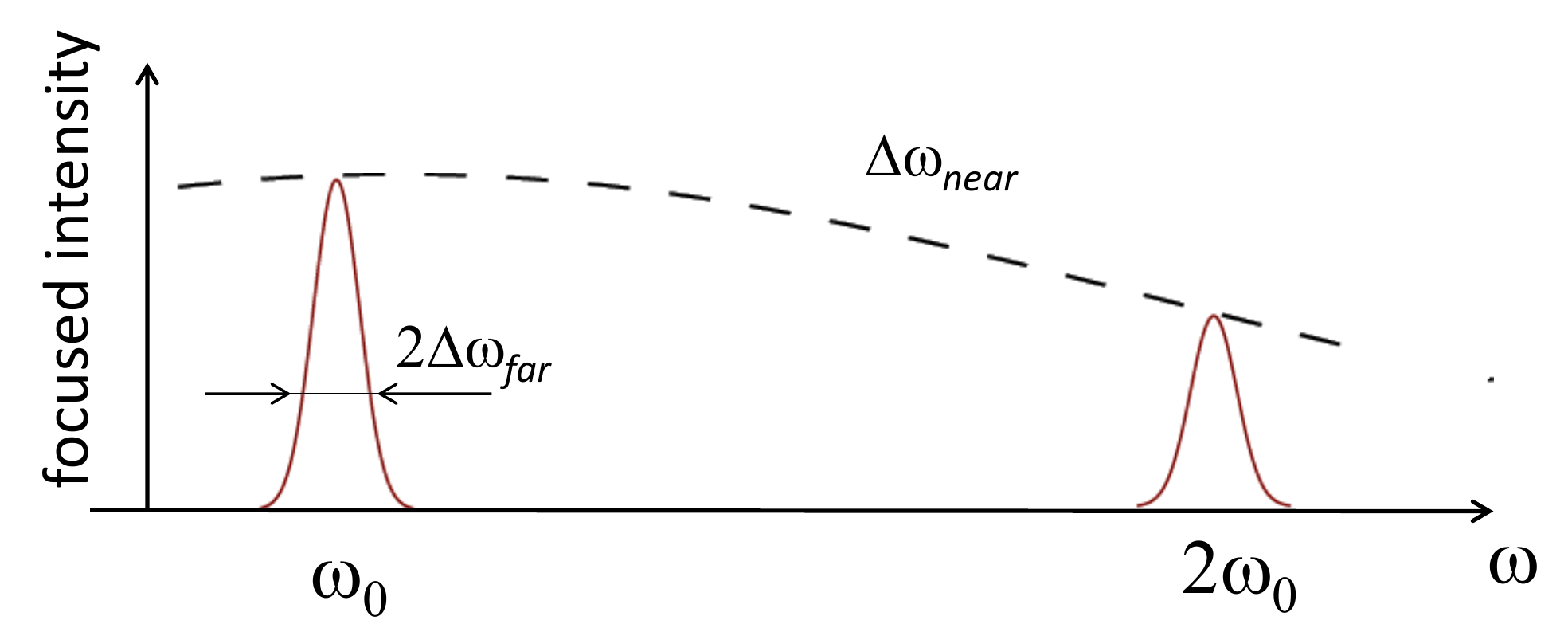}
\caption{(Color online) Focusing as a function of frequency in the
limit of low spatial dispersion.}
  \vskip -.1truein
\label{Fig-Summary}
\end{figure}

The bounds on the focusing imposed by the SLM are summarized in
Fig.\ref{Fig-Summary}. There can be several peaks of the focused
field, each of the width $2\D\w_{far}$, under the envelope of the
width $\D\w_{near}$.

Note that $\D\w_{near}$ is due to the pulse stretching at the
ISLM's pixel. If one moves the ISLM plane away from the surface of
the sample, then $\D\w_{near}$ approaches $\D\w_{far}$. Same
happens if the pixel size is increased, or if interference of
multiple pathways at each point of the ISLM plane becomes too
strong. We numerically verified that the peak at $2\w_0$ in
Fig.\ref{Fig-Summary} disappears if the ISLM pixels become so
large that they can not resolve the phase variations in the
scattered wave.

The peak also disappears if the  interference in the scattering
process can not be neglected.   In our calculations, this was
achieved by reducing the typical size of the impurities while
simultaneously increasing their number. This led to both higher
scattering angles, and deviations from the eikonal regime which
allows interpreting propagation of light via an ensemble of rays.
Surprisingly, however, predictions based on the eikonal optics
approach hold even for rather strong scattering: For light that
has passed through 30 3-$\mu$m thick planes with $\s=2$ $\mu$m
impurities, the phase compensation for the $\l_0=800$ nm far-field
focusing still provided a noticeable focal spot at $\l=\l_0/2=400$
nm.

Our consideration above refers to focusing laser pulses in space,
but not in time. Note, however that the ability to spatially focus
broadband light is bound by $\D\w_{far}$ - the spectral bandwidth
of a pulse which is not strongly distorted in the far field. Thus
we show that the above approach is limited to a spectral band
where the temporal structure is not destroyed, and temporal
focusing is not required. If the far-field stretching is
substantial, one needs to assign different pixels to different
bands, as described in the next Section. Only at that stage the
question of temporal shaping and focusing -- adjusting the
relative phases of several independent frequency bands -- arises.

\section{The general case      \label{Section-GeneralCase}}

In the general situation, many interfering pathways may lead to
the same near-field point. As before, let $\tnear$ characterize
the stretching of the pulse at the exit from the sample, and
$\tfar$ -- at infinity. Similar to the previous Section, the
near-field speckle pattern changes at detunings exceeding $\D\w_n
= \pi/\tnear$, and finite depth of the SLM's phase modulation
prevents focusing at detunings exceeding
$\D\w_f = \pi/\tfar$ except for the frequencies $\w_m$ related to $\w_0$ by %
\begin{equation}
\int k_m \,d S(k_m) = m \times \int k_0 \, dS(k_0) ~.
\end{equation} %
As shown in Section \ref{Section-MoreOngeometric}, if $\tnear <
\tfar$, it makes sense to place the ISLM in the near field with
respect to the random sample. Note that such situations include
those where the pulse is significantly modified after passing
through the random sample, both in space and in time.

In the case of stronger scattering, when the geometric optics
-based model is inapplicable, one must view the far-field focusing
as phasing together random phasors corresponding to different
scattering channels \cite{Mosk-OptLett07}. This regime is
mathematically similar to that of large ISLM pixels, discussed in
the previous Section (Eqs.(\ref{NearFieldSum-k0}-\ref{DwNear})).
Below we briefly discuss what scaling should be expected for
focusing broadband pulses in this case.

Let us assume that the far-field transmission spectrum within the
bandwidth of the laser pulse consists of $M$ un-correlated bands
of the width $\D\w_f$, and that the laser beam covers $N$ pixels
of the phase mask in the scheme of Fig.\ref{Fig-setup}(a). In
order to obtain figures of merit for the focusing capability, we
assign $N/M$ pixels of the SLM to each frequency band, in the way
that is discussed below. Assuming a  circular gaussian
distribution for the field amplitude after the sample
\cite{Mosk-OptLett07,Goodman-book},
\begin{equation}
P(E_{Re},E_{Im}) = \frac{1}{2\pi I_0} \exp  \left[ -\frac{E_{Re}^2
+ E_{Im}^2} {I_0} \right] \label{GaussianCircular}
\end{equation}
where $P$ is the probability density, and $E_{Re},E_{Im}$ are the
real and imaginary field amplitudes,  the intensity of the focused
field at frequency $\w$ is \cite{Goodman-book}
\begin{equation}
I_{coh}(\w) = \frac{\pi}{4}\left(\frac{N}{M}\right)^2 \,I_0 ~.
\label{GaussianCoherent}\end{equation}
In order for the focused spectrum to be controllable, this value
must exceed the background due to the other pixels assigned to all
other frequencies. The latter is obtained with the help of
Eq.(\ref{GaussianCircular}) as
\begin{equation}
\langle I_{back}(\w) \rangle = (M-1) \frac{N}{M} \,I_0 ~.
\end{equation}
Enhancement in spectral intensity due to the focusing is then
\cite{Mosk-OptLett07}
\begin{equation}
\eta_\w = \frac{\pi}{4} \frac{N}{M^2} ~. \label{EtaW}
\end{equation}
Once control over each spectral band of the width $\D\w_{far}$ is
achieved, one can tune the overall phase of the field in each band
by applying an extra phase to each phase mask's pixels controlling
the mode. Through these phases, the spatially focused pulse can
either be focused in time or be given any temporal shape allowed
by the frequency resolution of $\D \w$ and the number of pixels in
the phase mask. Thus the system makes an analog of a conventional
pulse shaper \cite{Wiener}, with the dispersion element being
replaced by the random sample
\cite{Silberberg-11,Lagendijk-Focusing-11}.

If the $M$ spectral components are given equal phase, together
they form a pulse that is $M$ times shorter in time than each of
the $M$ components. Its intensity is $M$ times higher than that of
the incoherent sum of the components. Thus the maximum achievable
intensity is
\begin{equation}
\eta_t = \frac{\pi}{4} \frac{N}{M} \label{EtaT}
\end{equation}
times stronger than that of un-compensated light.


\section{Summary    \label{Section-Conclusions}}

Coherent control of physical and chemical processes in turbid
media requires availability of focused laser pulses with tunable
temporal/spectral shapes. This, in turn, sets the task of
coherently controlling propagation of bi-chromatic and broadband
laser pulses through turbid media.

Recent works have shown that this task can be carried out by using
phase masks to adjust the phases of different transmission modes.
Each mode, centered at its own central frequency and having its
own speckle pattern, can coherently contribute to the output
field. By controlling the interference between the modes one can
achieve the desired spatio-temporal focusing. In this sense, the
experimental scheme shown in Fig.~\ref{Fig-setup} is an analog of
a conventional pulse shaper, with the dispersive element replaced
by the turbid sample. Resolution of this turbid pulse shaper is
set by the transmission properties of the sample
\cite{Silberberg-11}, together with one's technical ability to
control relative phases of the modes. The ability to shape pulses
simultaneously in space and time, and to work with very
narrow-band transmission modes, can bring new dimensions into
experiments on coherent control.

Most present-day experiments do not assume correlations between
the phase patterns of different frequency bands, and work in the
regime where the phase mask can not resolve phase variations
within a single speckle pattern. In this case one can obtain the
figure of merit for the efficiency of the spatio-temporal focusing
of light by assigning a fraction of the phase mask to each of the
independent frequency bands. This is done in
Section~\ref{Section-GeneralCase} of our paper
(Eqs.(\ref{EtaW},\ref{EtaT})). For $M$ independent frequency
bands, this leads to focused intensity at a single frequency
scaling as $\eta_\w\propto 1/M^2$. If the phases of the frequency
bands are set such as to produce a short pulse in the focus, its
intensity scales as $\eta_t\propto 1/M$.

 An interesting regime arises in the case of moderately
strong scattering and relatively large-size (above 2 $\mu$m in our
simulations with 800-nm light) scatterers. In this case, in
agreement with the geometric optics approach, optimization of
spatial focusing at frequency $\w_0$ automatically ensures that
focusing at the frequency $\w=n\w_0$ is also achieved. In this
situation, ''1+n'' coherent control must be available at no extra
cost provided the relative phase between the two fields can be
maintained. In addition, interesting phase structures arising at
frequencies that are rational fractions of $\w_0$ call for further
investigation.

In the geometric optics regime, the efficiency of the spatial
focusing is bound by the two time scales. A single set-up of the
phase mask can only optimize spatial focusing within a single
frequency transmission band, with the width given by $2
\D\w_{far}=2\pi / \tfar$ (Eq.\ref{FFbound}), where $\tfar$
corresponds to the stretching of an ultrashort pulse in the far
field at the output. At the same time, there is an overall
envelope of the focusing efficiency (Fig.\ref{Fig-Summary}). Its
width is given by $\D\w_{near} = \pi /\tnear$. Here $\tnear$ is
the duration of a pulse covered by the area of a single pixel of
the phase mask in the geometry of Fig.~\ref{Fig-setup}(b), and
$\D\w_{near}$ is the band with of the transmission matrix taken at
one pixel. The separation of the two time scales suggests that one
should choose the experimental set-up with the shortest $\tnear$.
To minimize the pulse stretching in the plane of the phase mask,
we proposed to image  the SLM onto the input surface of the turbid
sample.

The intuition inspired by geometric optics remains valid if one
considers far-field focusing at an angle, or if moderate
dispersion of the sample is taken into account. However, using the
same phase mask to focus at  frequencies $\w$ and $n\w$
simultaneously becomes difficult if the pixels of the phase mask
can not resolve individual near-field speckles. In this case $\D
\w_{near}$ approaches $\D\w_{far}$, and pulse stretching in the
near- and far field is the same. Then the geometry can not be
optimized by placing the phase mask at any particular distance
from the sample. Control over focusing of multiple frequencies can
be achieved by assigning subsets of the mask to different
frequency bands.}

\section*{Acknowledgements.}

The authors thank Moshe Shapiro, Azriel Z. Genack, and Stanislav
O. Konorov for valuable discussions. This work was supported by
DTRA, CFI and NSERC. E.S. acknowledges the Institute of
Theoretical Atomic, Molecular, and Optical Physics (ITAMP) for
support during a visit to ITAMP facilities.

\end{document}